\documentclass[aps]{revtex4}
\renewcommand\ln{\log}
\newcommand\ba{\begin{eqnarray}}
\newcommand\ea{\end{eqnarray}}

\usepackage{graphicx}
\usepackage{dcolumn}
\usepackage{bm}

\begin{document}


\title{
Electroweak corrections to the Drell-Yan process
in the high dimuon mass range}

\author{Vladimir A. Zykunov}
\email{zykunov@sunse.jinr.ru, zykunov@gstu.gomel.by}
\affiliation{
Joint Institute for Nuclear Research, 141980, Dubna, Russia
and\\
Gomel State Technical University, 246746, Gomel, Belarus
}%

\date{\today}

\begin{abstract}
The complete electroweak radiative ${\cal O}{(\alpha)}$ 
corrections to the Drell-Yan process
at large invariant dimuon mass have been studied.
All formulas for the cross sections and kinematical restrictions
are presented in explicit form, for the simplification of calculation 
and coding the $\theta$-- and $\delta$--functions are actively used.
The FORTRAN code READY for the numerical analysis  
in the high energy region corresponding to the future experiments
at the CERN Large Hadron Collider has been constructed.
To simulate the detector acceptance we used the standard CMS detector cuts.
The radiative corrections are found to become large at 
high dimuon mass $M$, the complete corrections at "bare" setup change 
the dimuon mass distribution up to $\sim -5.6\%\ (-23.2\%;\ -35.3\%) $
at the LHC energy and $M=1\ (3;\ 5) \mbox{TeV}$.
\end{abstract}


\maketitle

\section{Introduction}
More than twenty years the Standard Model (SM) keeps for oneself 
the status of consistent and experimentally confirmed theory since 
the experimental data of past and present accelerators (LEP, SLC and Tevatron)
have shown no significant deviation from the SM predictions 
up to energy scales of several hundred GeV. 
However, New Physics (NP) models: 
various left-right symmetric models, extended gauge theories including 
grand unification theories, models of composite gauge bosons 
\cite{34},
some extra dimension scenarios \cite{extra-dim},
extra neutral gauge bosons \cite{extra-bos} 
and fermion compositeness models 
\cite{89}
predict various deviations out of 
the SM predictions and, therefore, their testing in the new energy scale 
(thousands GeV) is one of the main tasks of modern physics. 
The forthcoming experiments at the collider LHC 
provides such possibility and probably will shed the light on this 
important problem in the immediate future.

The experimental investigation of the continuum
for the Drell-Yan production of a dilepton pair, i.e.
data on the cross section and  the forward-backward asymmetry of
the reaction
\begin{equation}
pp \rightarrow \gamma,Z  \rightarrow \mu^+\mu^-X
\label{1}
\end{equation}
at large invariant mass of a dimuon pair (see \cite{cmsnote}
and references therein)
is considered to be one of the powerful tool in the experiments 
at the LHC from the NP exploration standpoint.

\begin{figure*}
\vspace*{15mm}
\hspace*{-5mm}
\scalebox{0.25}{\includegraphics{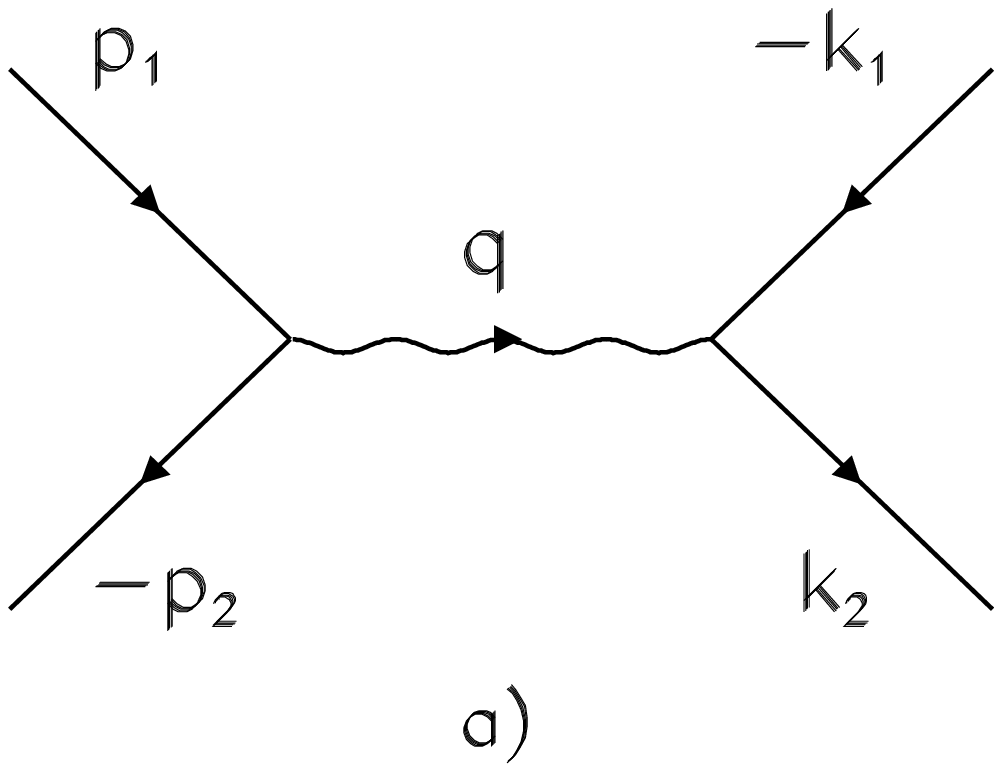}}
\scalebox{0.25}{\includegraphics{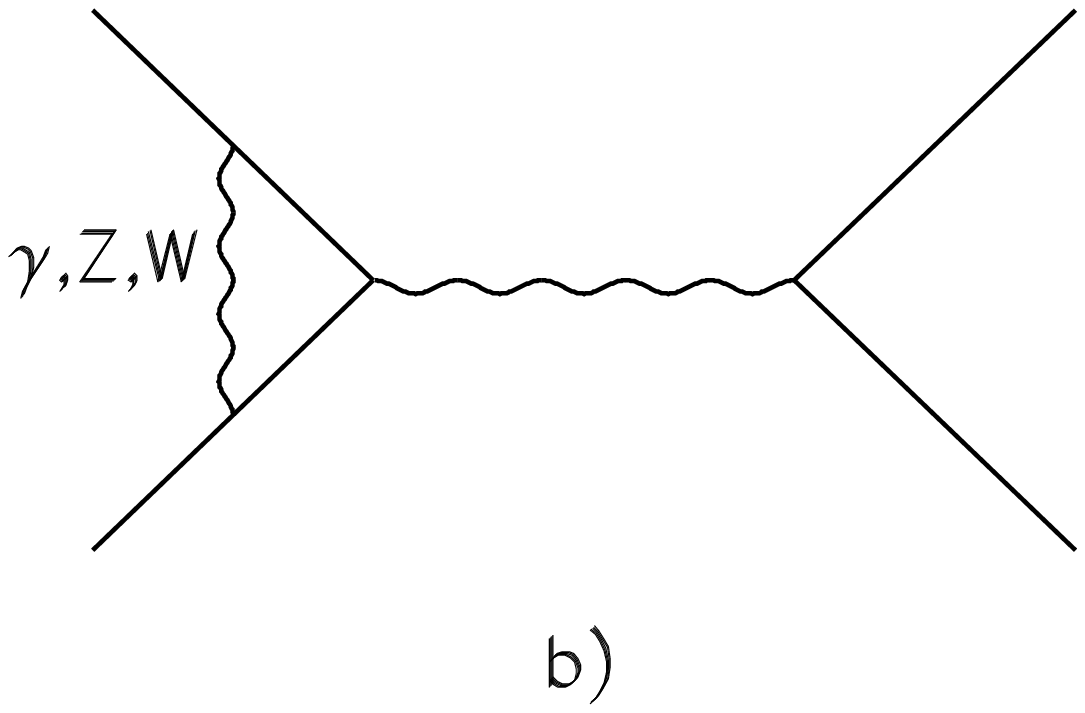}}
\scalebox{0.25}{\includegraphics{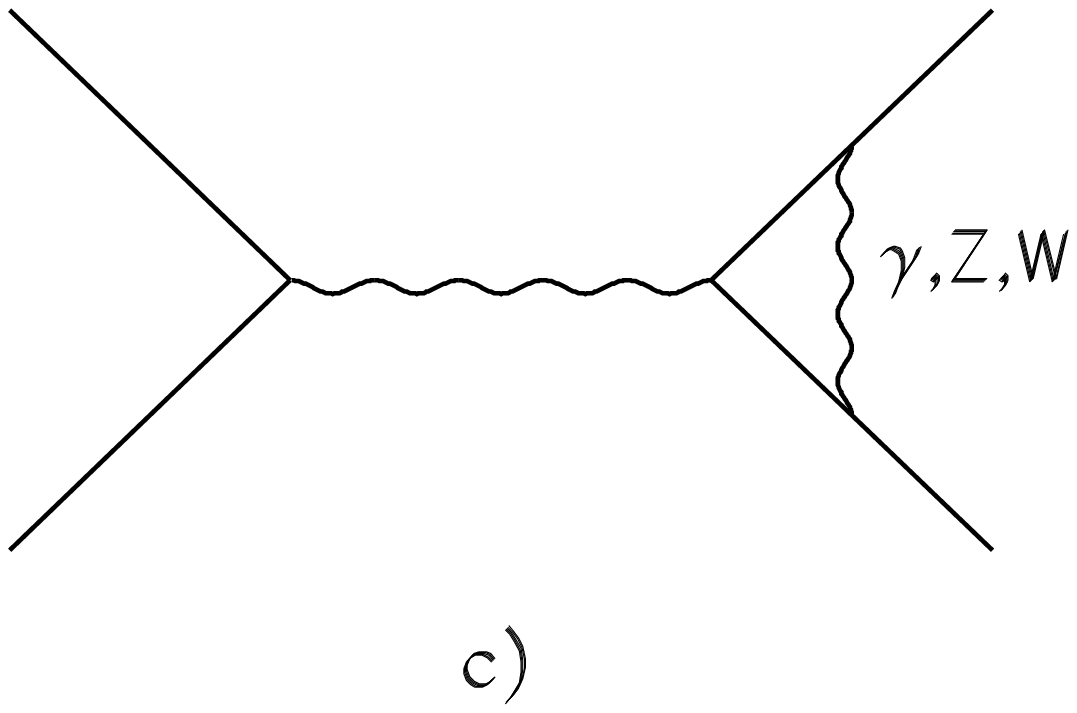}}
\scalebox{0.25}{\includegraphics{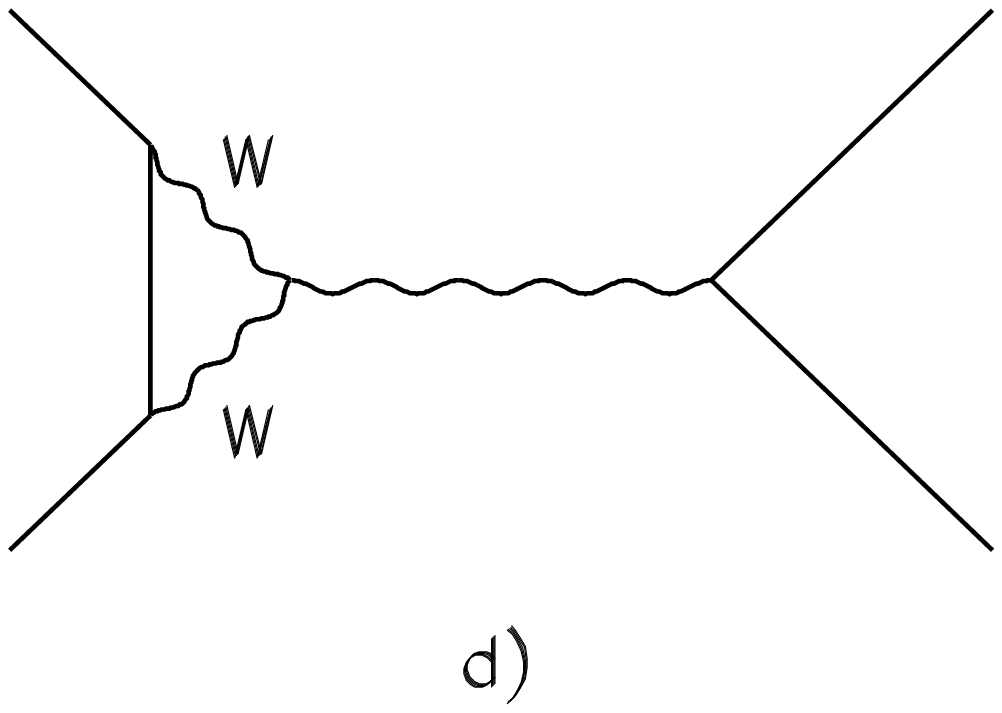}}
\scalebox{0.25}{\includegraphics{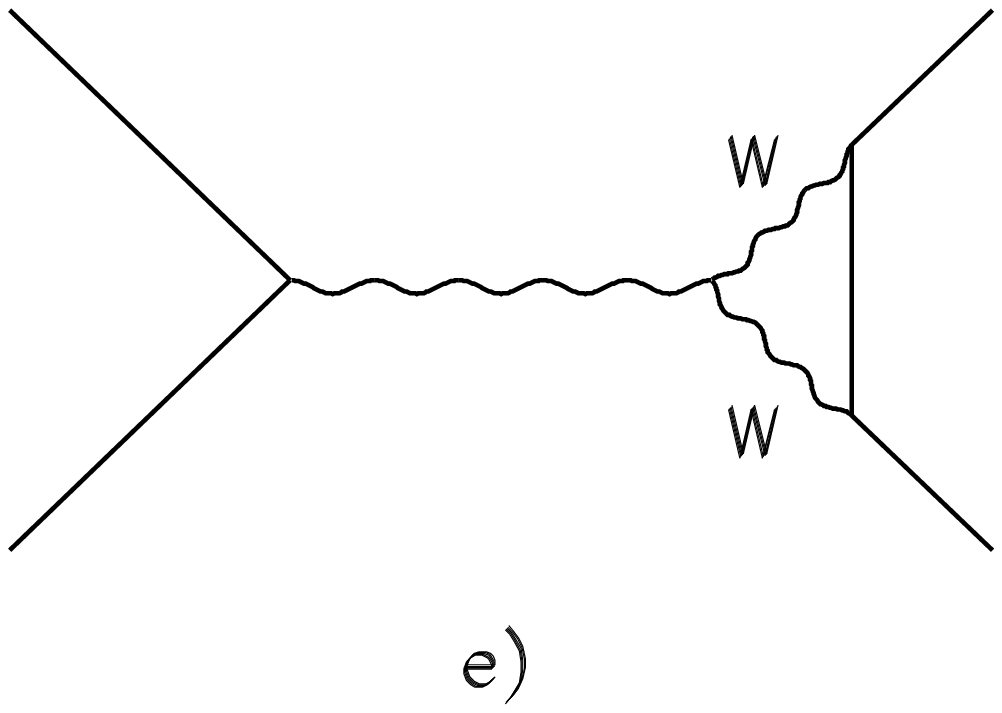}}
\hspace*{6mm}
\scalebox{0.25}{\includegraphics{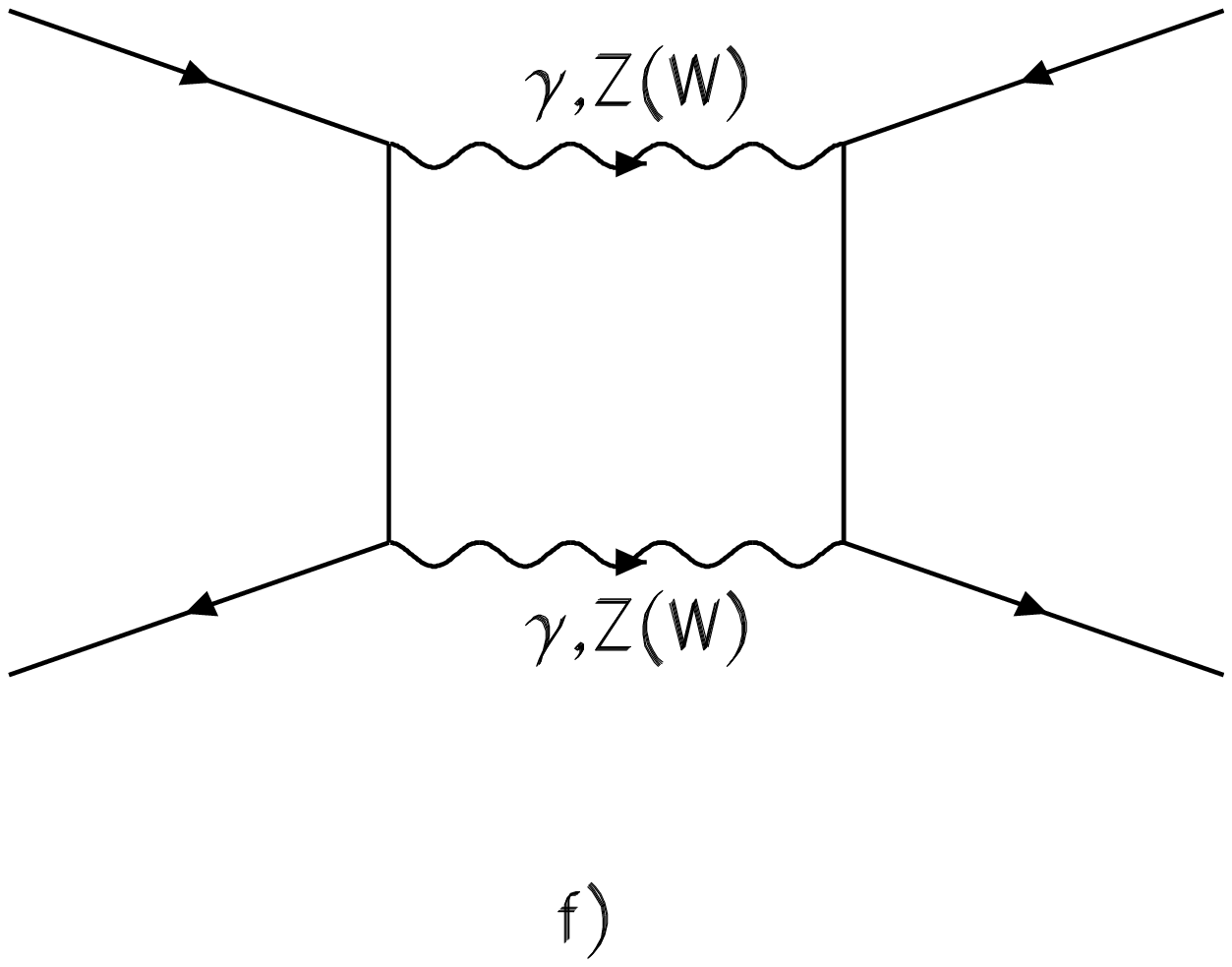}}
\hspace*{7mm}
\scalebox{0.25}{\includegraphics{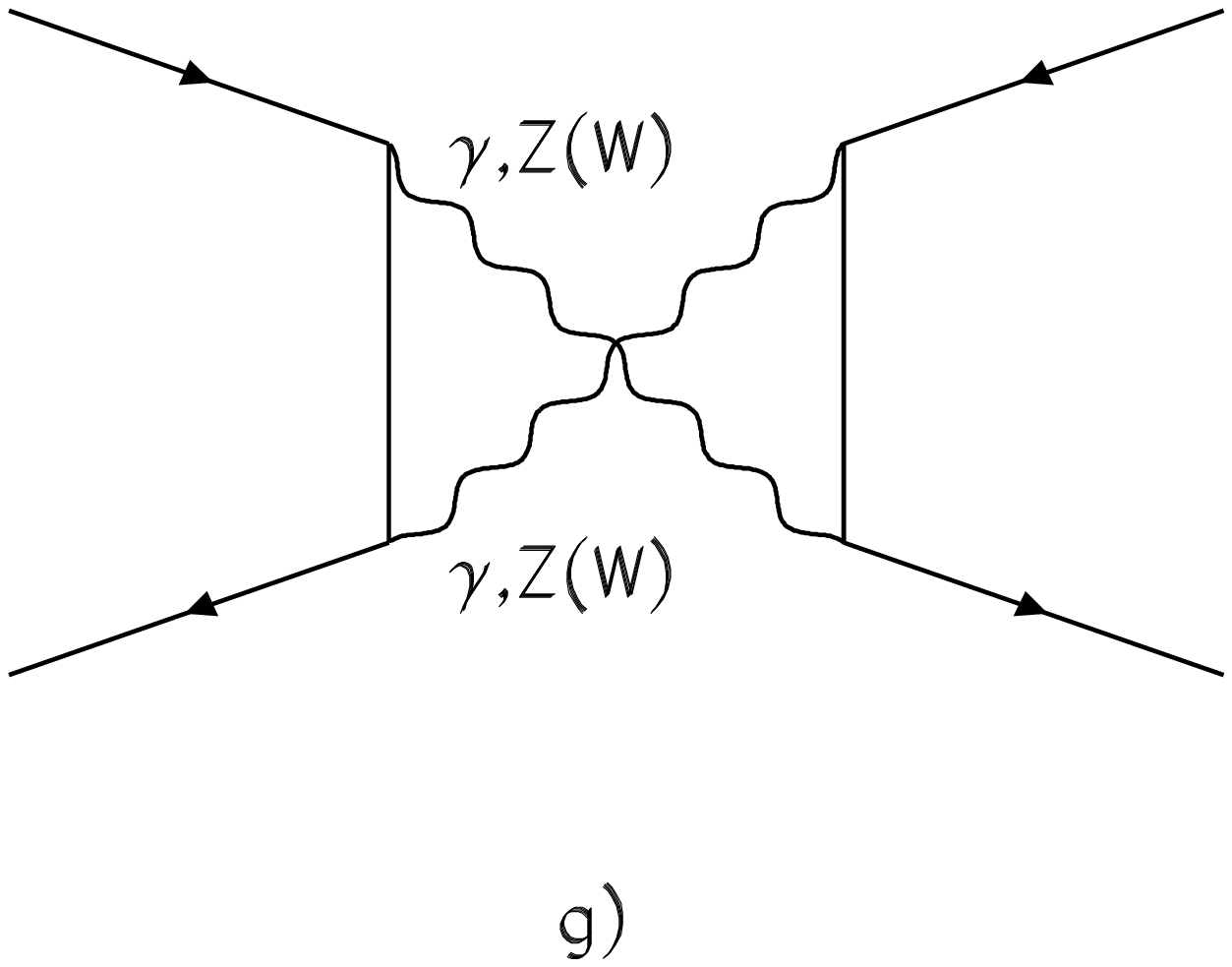}}
\hspace*{10mm}
\scalebox{0.25}{\includegraphics{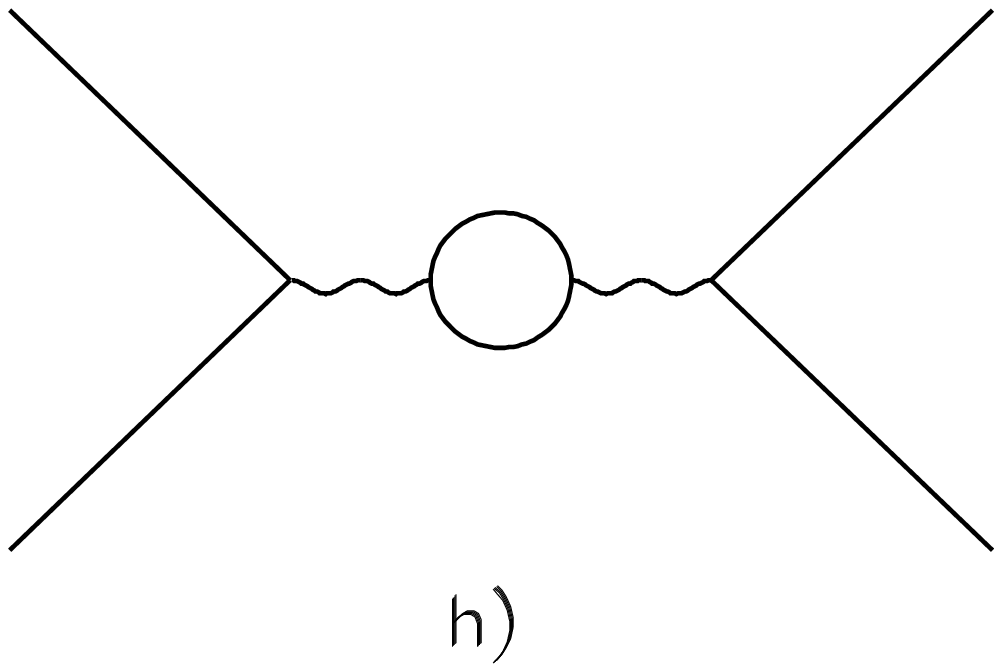}}
\hspace*{20mm}
\scalebox{0.25}{\includegraphics{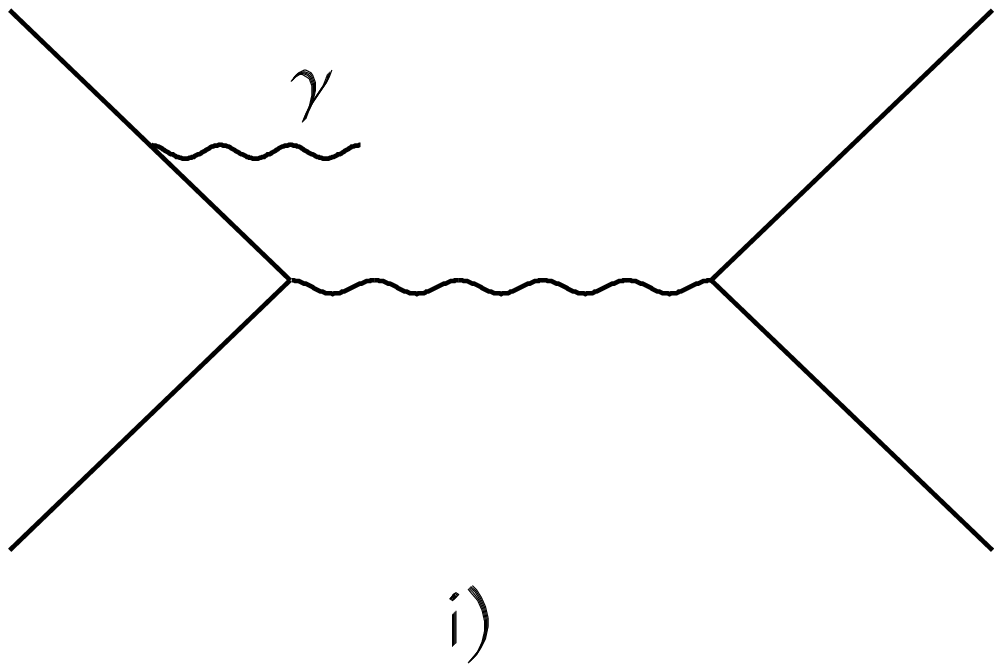}}
\scalebox{0.25}{\includegraphics{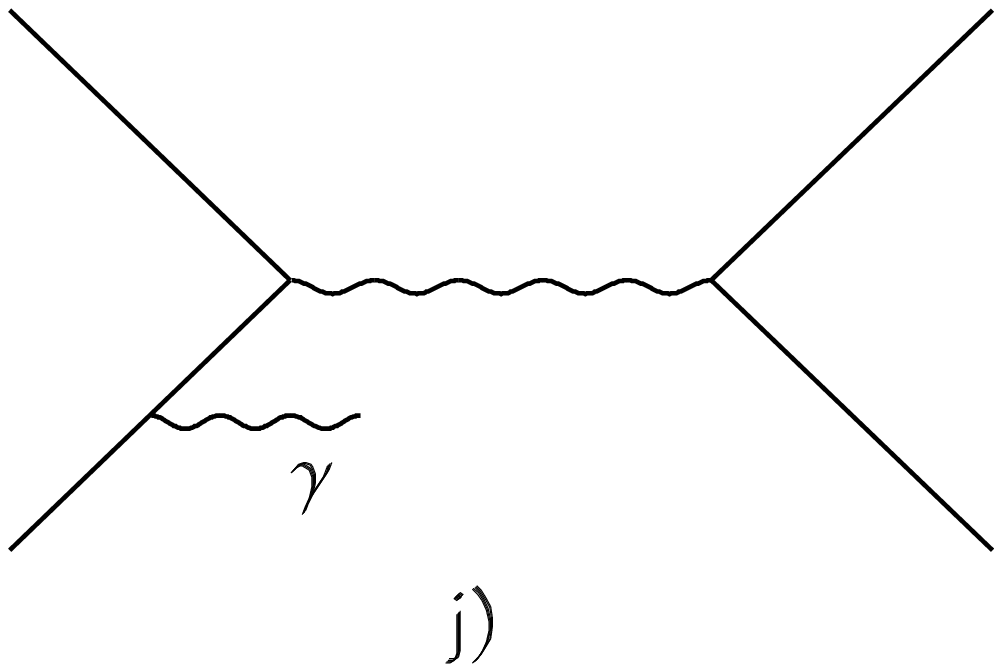}}
\scalebox{0.25}{\includegraphics{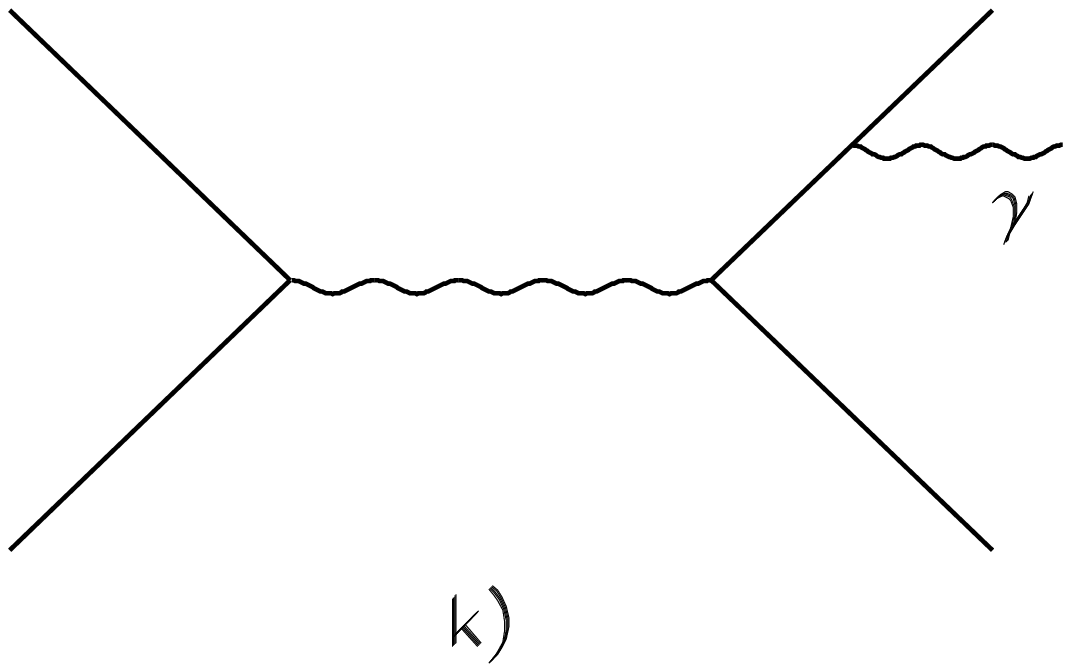}}
\scalebox{0.25}{\includegraphics{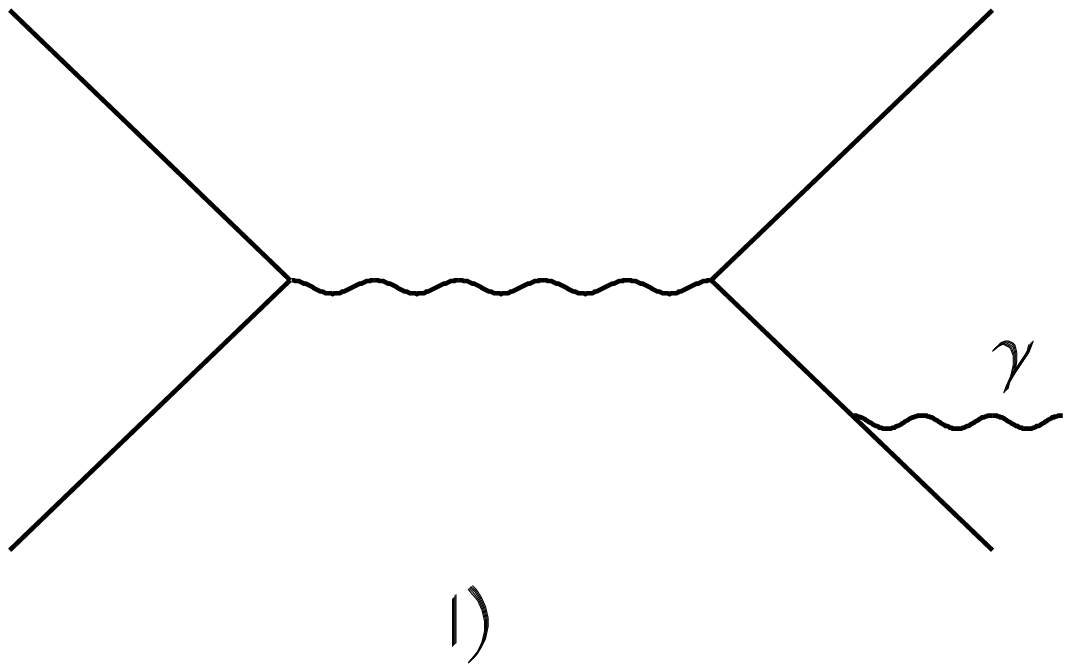}}
\vspace*{-25mm}
\caption{\label{fig:1}
Feynman graphs for the Born (a), one-loop virtual diagrams (b-h) 
and bremsstrahlung diagrams (i-l).
Unsigned helix lines mean $\gamma$ or $Z$.
}
\end{figure*}

The studies of the NP effects are impossible without the
exact knowledge of the SM predictions including higher order
QCD and ElectroWeak radiative Corrections (EWC).
The EWC to the reaction (\ref{1}) are studied well (see papers on
pure QED corrections \cite{MosShuSor}, and  the QED and electroweak
corrections in the
Z-peak region and above in \cite{DY2002}
and numerous papers cited there).
The EWC result contains so-called Double Sudakov Logarithms (DSL) 
\cite{sud-log},
i.e. the expressions which are growing with the scale of energy,
and thus giving one of the main effect in the region of large
invariant dimuon mass.
By now extensive studies have been done in this area.  
For instance, the weak Sudakov expansion for general 
four-fermion processes has been studied in detail 
(see, for example, \cite{DENPOZ} 
and the recent paper \cite{ARX05} along with 
the extensive list of references therein).
Obviously, the collinear logarithms of QED radiative
corrections can compete with the DSL in the investigated region.
This important issue has been studied
at the one-loop level in \cite{DY2002},
where both the QED and weak corrections have been calculated
for $M \leq$~2~TeV,
but it has yet remained unsolved in the region of $M>$~2~TeV
(with the exception of short numerical estimation of EWC 
to $pp \rightarrow e^+e^-X$ by program ZGRAD \cite{ZGRAD} in \cite{Baur2006}, 
see Fig.7 there).
Other important contributions in the investigated reaction
at high invariant masses
are the higher-order corrections (two-loop electroweak logarithms, at least),
which also have been studied in the works 
 \cite{ARX05}, \cite{Denner:2006jr}
(see also the numerous papers cited therein).
Weak boson emission contribution has been recently calculated 
in \cite{Baur2006}
and the contribution of higher-order corrections due to multiple photon 
emission has been computed in \cite{CarCal},
these contributions are beyond the presented calculations.

Thus, for the future experiments at LHC
aimed at the searches of NP in the reaction (\ref{1})
it is urgent to know exactly the SM predictions,
including the radiative background, i.e. the processes,
which are experimentally indistinguishable from (\ref{1}).
The important task is the insertion of this background into the 
LHC Monte Carlo generators and they should be both accurate 
and fast. For them to be fast it is necessary to have a set of 
compact formulas for the EWC. They are obtained in our previous paper
\cite{YAFDY} using the Asymptotic Approach (AA) and improved
in subsequent paper \cite{PRD}.
To speed up the bremsstrahlung calculation and increase its 
accuracy it is very important to have all formulas 
for the cross sections and kinematical restrictions in explicit form
appropriate for adaptive multidimensional integration.
We obtained such form here and reached the high speed 
and good accuracy of calculation using the $\theta$-- and $\delta$--functions 
apparatus.

\section{Notations and the cross section with the Born kinematics}
Our notations are the following (see Fig.1,a):
$p_1 (p_2)$ is the 4-momentum of the first (second) quark 
or antiquark with the flavor $q$ and mass $m_q$;\
$k_1 (k_2)$ is the 4-momentum of the final muon $\mu^+ (\mu^-)$
with the mass $m$;\ \
$q=k_1+k_2$ is the 4-momentum of the $i$-boson with the mass $m_i$
($i=\gamma, Z$);\ \
$P_{A(B)}$ is the 4-momentum of initial nucleon $A(B)$.
We use the standard set of Mandelstam invariants for the partonic elastic
scattering $s,\ t,\ u$:
\begin{equation}
s=(p_1+p_2)^2,\ t=(p_1-k_1)^2,\ u=(k_1-p_2)^2,
\end{equation}
and the invariant \ $S=(P_A+P_B)^2$  for hadron scattering.
The invariant mass of dimuon is $M=\sqrt{q^2}$.

For a start let us present the convolution formula for
the total hadronic (H) cross section, where we used such abbreviations
and indices:
"Born"\ (index 0), 
V-contribution: (indices: 
BSE for boson self energies, 
HV for "heavy" vertices,
"$\gamma \gamma$" for the IR-finite part of $\gamma \gamma$-boxes,
"$\gamma Z$" for the IR-finite part of $\gamma Z$-boxes,
"$ZZ$" for the $ZZ$-boxes,
"$WW$" for the $WW$-boxes,
"$fin$" for the sum of "Light" Vertices (LV), 
   the IR-part of $\gamma\gamma$ and $\gamma Z$--boxes
   and emission of "soft" photon, with the energy less than $\omega$).
The "$fin$"-part is IR-finite and described by Born kinematics. 
Also we used common index for V-contribution
$V=0,\mbox{BSE},\mbox{HV},b,fin$ 
and special index for boxes  $b=\gamma\gamma,\gamma Z,ZZ,WW$.
Thus, the hadronic cross section looks like
\begin{eqnarray}
\sigma_{V}^H =
\frac{1}{3}
\int\limits_{0}^{1}dx_1
\int\limits_{0}^{1}dx_2
\int\limits_{-S}^{0}dt
\sum\limits_{q=u,d,...}
&&[ f_q^A(x_1,Q^2)f_{\bar q}^B(x_2,Q^2) \hat \sigma_{V}^{q\bar q}(t) +
f_{\bar q}^A(x_1,Q^2)f_{q}^B(x_2,Q^2) \hat \sigma_{V}^{\bar q q}(t)]
\theta(t+\hat s)
\theta_M
\hat \theta_{D},
\label{xsfin}
\end{eqnarray}
here
the $f_q^H(x,Q^2)$ is the probability at energy scale $Q^2$ 
of constituent $q$ 
with the fraction $x$ of the hadron's momentum in hadron $H$
finding,
$\hat s=x_1 x_2 S$, 
the $\theta$-function under integral sign determined by
the kinematics of parton reaction,
the factor 
\begin{equation}
\theta_M=\theta(\hat s-M_1^2)\theta(M_2^2-\hat s)
\end{equation}
provides the integration in interval of invariant  mass
$M_1 \leq M \leq M_2$ and 
the factor
\begin{equation}
\theta_{D}=
\theta(\zeta^*-\cos\theta)\theta(\zeta^*+\cos\theta)
\theta(\zeta^*-\cos\alpha)\theta(\zeta^*+\cos\alpha)
\theta(p_T(\mu^+)-p_T^{min})\theta(p_T(\mu^-)-p_T^{min})
\label{tetad}
\end{equation}
cuts the region of integration according 
detector geometry, 
the parameters $\zeta^*$ and $p_T^{min}$ will be discussed below.
The expressions for the angles 
($\theta\ (\alpha)$ is the scattering angle of the muon
with the 4-momenta $k_1\ (k_2)$ in the center mass system of hadrons)
and energies (also in the centre of hadron mass system)
can be obtained as special situation of radiation case 
("radiative" invariants $v,\ z,\ u_1,\ z_1$=0) from the formulas 
(\ref{uie}) presented below.
For transverse components the expressions take place:
$p_T(\mu^+)={k_1}_0 \sin\theta,\
p_T(\mu^-)={k_2}_0 \sin\alpha$.

Let us enumerate all quark cross sections in (\ref{xsfin})
using agreement $\sigma(t) \equiv d\sigma/{dt}$.

The Born cross section looks like
\begin{eqnarray}
\sigma_{0}^{q\bar q}(t) =
\frac{2\pi\alpha^2}{s^2}
\sum\limits_{i,j=\gamma,Z} D^{i}{D^{j}}^* (b_+^{i,j}t^2+b_-^{i,j}u^2),
\end{eqnarray}
where
the non-radiative boson propagators look like
\begin{equation}
D^{j} =\frac{1}{s-m_j^2+im_j\Gamma_j},
\end{equation}
$\Gamma_j$ is the $j$-boson width,
\begin{equation}
b_{\pm}^{n,k}=
{\lambda_q}^{n,k}_+{\lambda_l}^{n,k}_+
\pm {\lambda_q}^{n,k}_-{\lambda_l}^{n,k}_-
\label{bplusminus}
\end{equation}
and 
the combinations of coupling constants for $f$-fermion with $i$- (or $j$-)
boson have the form
\begin{equation}
{\lambda_f}^{i,j}_+=v^i_fv^j_f+a^i_fa^j_f,\
{\lambda_f}^{i,j}_-=v^i_fa^j_f+a^i_fv^j_f,
\label{lamb}
\end{equation}
where
\begin{equation}
  v^{\gamma}_f=-Q_f,\
  a^{\gamma}_f=0,\
  v^Z_f=\frac{I_f^3-2s_W^2Q_f}
             {2s_Wc_W},\
  a^Z_f=\frac{I_f^3}{2s_Wc_W},
\end{equation}
$Q_f$ is the electric charge of fermion $f$,
$I_f^3$ is the third component of the weak isospin of fermion $f$, and
$s_W\ (c_W)$ is the sine(cosine) of the weak mixing angle:
$s_W=\sqrt{1-c_W^2}$, $c_W= m_W/m_Z$.

The BSE-part is
\begin{eqnarray}
\sigma^{q \bar q}_{\rm BSE}(t)=-\frac{4\alpha^2\pi}{s^2} \bigl[ &&
\sum\limits_{i,j=\gamma,Z}
\Pi_S^i D^{i} {D^{j}}^* \sum\limits_{\chi=+,-}
{\lambda_q}^{i,j}_{\chi} {\lambda_l}^{i,j}_{\chi} B_{\chi} +
\nonumber \\&&
+  \Pi_S^{\gamma Z} D^{Z} \sum\limits_{i=\gamma,Z}
{D^{j}}^* \sum\limits_{\chi=+,-}
( {\lambda_q}^{\gamma,j}_{\chi} {\lambda_l}^{Z,j}_{\chi} +
{\lambda_q}^{Z,j}_{\chi} {\lambda_l}^{\gamma,j}_{\chi}) B_{\chi} \bigr].
\end{eqnarray}
Here
$\Pi_S^{\gamma,Z,\gamma Z}$ are connected with 
the renormalized photon--, Z-- and $\gamma$Z--self energies 
\cite{BSH86,Hollik} as
$$ 
\Pi_S^{\gamma}=\frac{\hat\Sigma^{\gamma}}{s},\
\Pi_S^{Z}=\frac{\hat\Sigma^{Z}}{s-m_Z^2},\
\Pi_S^{\gamma Z}=\frac{\hat\Sigma^{\gamma Z}}{s}.$$

The HV-part looks like
\begin{eqnarray}
\sigma^{q \bar q}_{\rm HV}(t)=
   \frac{4 \pi \alpha^2}{s^2}
   {\rm Re}
   \sum_{i,j=\gamma,Z} D^{i} {D^{j}}^*
   \sum_{\chi=+,-} 
({\lambda_q^{\rm F}}^{i,j}_{\chi} {\lambda_l}^{i,j}_{\chi} +
 {\lambda_q}^{i,j}_{\chi} {\lambda_l^{\rm F}}^{i,j}_{\chi}) B_{\chi},
\end{eqnarray}
where the form factors ${\lambda_f^{\rm F}}^{i,j}_{\pm}$
are explained in \cite{YAFDY}.

The boxes can be presented as
\begin{eqnarray}
\sigma^{q \bar q}_{ b}(t)=
   \frac{2 \alpha^3}{s^2}
   \sum_{k=\gamma,Z} {D^{k}}^*
(\delta^{b,k}(t,u,b_+,b_-)-\delta^{b,k}(u,t,b_-,b_+)),
\end{eqnarray}
where functions $\delta^{b,k}(t,u,b_+,b_-)$
and all prescriptions for them can be found in \cite{YAFDY,PRD}.

The "$fin$"-part (the result of infrared singularity cancellation of 
$\gamma\gamma, \gamma Z, \mbox{LV}$ and "soft" bremsstrahlung) is
\begin{eqnarray}
\sigma_{fin}^{q\bar q}(t) = &&
\frac{\alpha}{\pi} 
\delta_{fin}^{q\bar q}
\sigma_{0}^{q\bar q}(t), 
\nonumber \\[0.3cm] \displaystyle
\delta_{fin}^{q\bar q} =&&
J_0 \ln\frac{2\omega}{\sqrt{s}}
+ Q_l^2 (\frac 32 \ln\frac{s}{m^2}-2+\frac{\pi^2}{3})
+ Q_q^2 (\frac 32 \ln\frac{s}{m_q^2}-2+\frac{\pi^2}{3})
\nonumber \\[0.3cm] \displaystyle
&&-Q_qQ_l (\ln\frac{s^2}{tu}\ln\frac{t}{u}+\frac{\pi^2}{3}
+\ln^2\frac{t}{u}+4{\mbox Li}_2\frac{-t}{u}), 
\nonumber \\[0.3cm] \displaystyle
&&J_0=2\bigl(Q_q^2\bigl(\ln\frac{s}{m_q^2}-1\bigr)-2Q_qQ_l\ln\frac{t}{u}
+Q_l^2\bigl(\ln\frac{s}{m^2}-1\bigr)\bigr),
\label{soft}
\end{eqnarray}
where $\mbox{Li}_2$ denotes the Spence dilogarithm.
Let us note that correction $\delta_{fin,{\rm FSR}}^{q\bar q}$ 
is well known and presented, for example, in paper \cite{baur},
and the correction $\delta_{fin,{\rm ISR}}^{q\bar q}$ can be found
in the following way
$$\delta_{fin,{\rm ISR}}^{q\bar q} = \delta_{fin,{\rm FSR}}^{q\bar q}
(m\rightarrow m_q, Q_l\rightarrow Q_q).$$

To find the cross section for ${\bar q q}$-case, it is necessary to
change $t \leftrightarrow u$ in the Born part 
and $Q_qQ_l \rightarrow -Q_qQ_l$ in "$fin$"-part. 
The "hat" in formula (\ref{xsfin}) means only $s \rightarrow \hat s$.

\section{"Hard"\ photons}

Let us present the hadronic cross section induced by bremsstrahlung
(Fig.1,i-Fig.1,l).
Introducing
the total phase space of reaction as
\begin{eqnarray}
I_{\Omega}^6[A]=
\int\limits_{0}^{1}dx_1
\int\limits_{0}^{1}dx_2
\int \!\!\!\! \int\limits_{~\Omega} \!\!\!\! \int \!\!\!\! \int dt dv dz du_1
\frac{1}{\pi\sqrt{R_{u_1}}} 
\theta(\hat R_{u_1}) 
\theta_M^R
\hat \theta_{D}^R \ A,
\label{i6}
\end{eqnarray}
where $z=2k_1p,\ v=2k_2p,\ z_1=2p_1p,\ u_1=2p_2p$ and
$p$ is the 4-momentum of a real bremsstrahlung photon.

The factors $\theta_M^R$ and $\theta_{D}^R$ look in "radiative"\ case
slightly different in comparison with "non-radiative" ones:
\begin{equation}
\theta_M^R=\theta(\hat s-z-v-M_1^2)\theta(M_2^2-\hat s+z+v),
\end{equation}
and for $\theta_{D}^R$ we use "non-radiative"\
expression $\theta_{D}$ (\ref{tetad}),
we should change only the angles and energies:
\begin{eqnarray}
&&\cos\theta=1+\frac{t}{{k_1}_0 x_1 \sqrt{S}},\ 
\cos\alpha=1+\frac{u+z-u_1}{{k_2}_0 x_1 \sqrt{S}},
\nonumber \\[0.3cm] \displaystyle
&&{k_1}_0=-\frac{1}{2\sqrt{S}} \bigl(\frac{t}{x_1}+\frac{u}{x_2}\bigr),\
{k_2}_0=  \frac{1}{2\sqrt{S}}
\bigl( \frac{s+t-z_1}{x_1}+\frac{s+u-u_1}{x_2} \bigr), 
\label{uie}
\end{eqnarray}
where $u=v-\hat s-t$ and $z_1=z-u_1+v$.

The physical region $\Omega$ is determined by 
$\theta(R_{u_1})$, where $R_{u_1}$
($R_{u_1}$ is the Gram determinant multiplied by constant factor)
is described by:
\begin{eqnarray}
R_{u_1}&=& -A_{u_1} u_1^2 - 2 B_{u_1} u_1 - C_{u_1},
\nonumber \\[0.3cm] \displaystyle
A_{u_1}&=& -4m^2s + (s-v)^2,
\nonumber \\[0.3cm] \displaystyle
B_{u_1}&=& v[ m^2 (3 s - v) + (s - v) (m_q^2 - s - t + v) ]+
           z[  m^2(s - v) - m_q^2(s + v) + st + v(s+t-v) ],
\nonumber \\[0.3cm] \displaystyle
C_{u_1}&=& z^2[ (m^4 + m_q^4 - 2 m^2(m_q^2 + t - v)
               - 2 m_q^2 ( t + v) + (t - v)^2 ] +
\nonumber \\[0.3cm] \displaystyle
  &+& 2zv [
m^4 + m_q^4 + m_q^2(s - 2 t) - 
m^2 (2 m_q^2 + s + 2 t - 2 v) + (t - v)(s + t - v)
] +
\nonumber \\[0.3cm] \displaystyle
  &+& v^2 [ m^4 - 2 m^2 (m_q^2 + s + t - v) + (m_q^2 - s - t + v)^2 ].
\end{eqnarray}

Then the total bremsstrahlung cross section have form:
\begin{eqnarray}
\sigma_{R}^H =&&
\frac{\alpha^3}{3}
I_{\Omega}^6 [ \ \hat s^{-2} \!\!\!
\sum\limits_{\chi=+,-} \
\sum\limits_{q=u,d,...}
\sum\limits_{i,j=\gamma,Z} 
{\lambda_q}^{i,j}_{\chi} {\lambda_l}^{i,j}_{\chi} \times 
\nonumber \\[0.3cm] \displaystyle
&&
\bigl(
[ f_q^A(x_1,Q^2)f_{\bar q}^B(x_2,Q^2) + \chi
  f_{\bar q}^A(x_1,Q^2)f_{q}^B(x_2,Q^2) ]
[ Q_l^2 {R_l}_{\chi}^{q\bar q} D^{i}{D^{j}}^* 
 +Q_q^2 {R_{qk}}_{\chi}^{q\bar q} \Pi^{i}{\Pi^{j}}^* ]
\nonumber \\[0.3cm] \displaystyle
&& \left.
+ [ f_q^A(x_1,Q^2)f_{\bar q}^B(x_2,Q^2) - \chi
  f_{\bar q}^A(x_1,Q^2)f_{q}^B(x_2,Q^2) ]
 Q_lQ_q {R_{int}}_{\chi}^{q\bar q} 
\frac{\Pi^{i}{D^{j}}^* + D^{i}{\Pi^{j}}^*}{2}
\bigr)\right|_{s \rightarrow \hat s} ].
\label{xshard}
\end{eqnarray}
Indices $l,\ qk$ and $int$ mean the origin of emmited photon:
lepton, quark and lepton-quark interference, i.e.
Final State Radiation (FSR), Initial State Radiation (ISR) 
and their INTerference (INT), correspondingly.
The "radiative" boson propagators look like
\begin{equation}
\Pi^{j} =\frac{1}{s-z-v-m_j^2+im_j\Gamma_j}
\end{equation}
and the expressions $R$ can be found in Appendix A. 

Dissecting the region of integration with the help of function
$$\theta_{\omega} = \theta(\frac{v+z}{2\sqrt{s}}-\omega),$$
we divide the cross section (\ref{xshard}) in two parts:
first one is corresponding to "soft"\ photons with the energy less then 
$\omega$ (it goes to IR singularity cancellation in formula (\ref{soft}) 
of Sect.I) and the second one is corresponding to "hard"\ photons
with the energy more then $\omega$. 
We realize the numerical integration of (\ref{i6}) 
(and, certainly, of (\ref{xsfin})) 
by Monte Carlo routine based on the VEGAS algorithm \cite{VEGAS}.

\section{Discussion of numerical results}
\begin{figure*}
\vspace*{45mm}
\hspace*{-30mm}
\scalebox{0.3}{\includegraphics{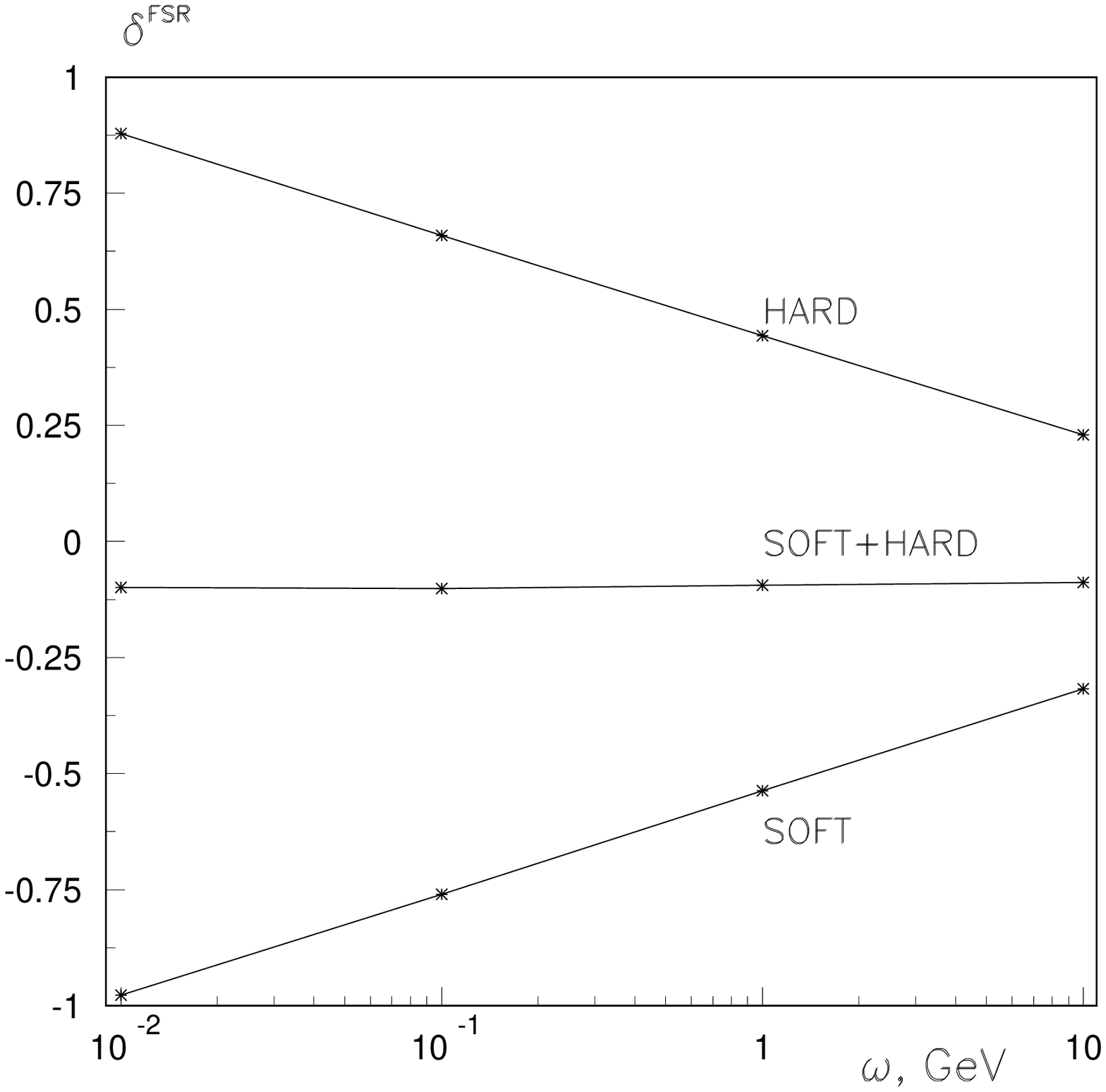}}
\hspace*{25mm}
\scalebox{0.3}{\includegraphics{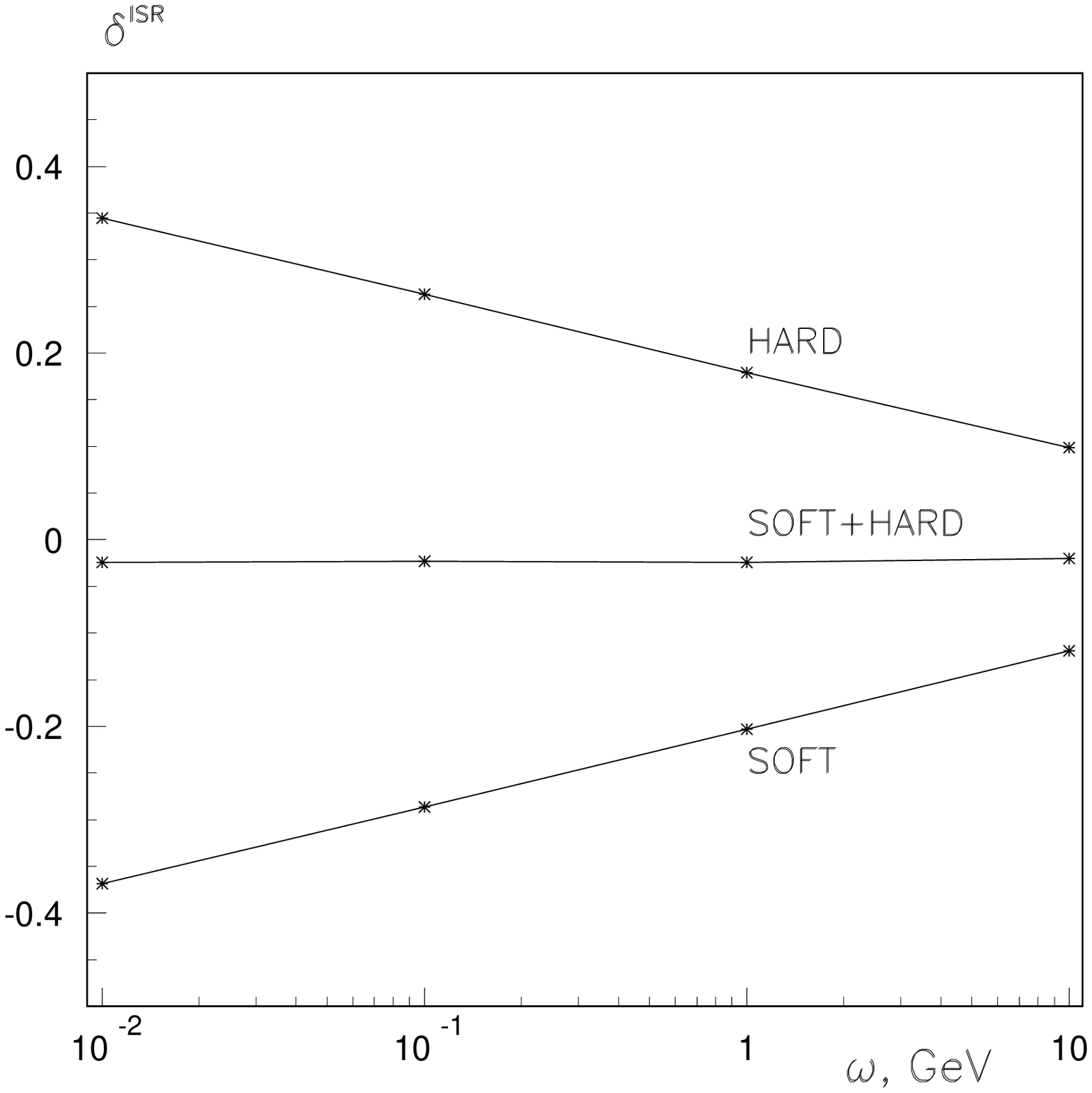}}
\hspace*{25mm}
\scalebox{0.3}{\includegraphics{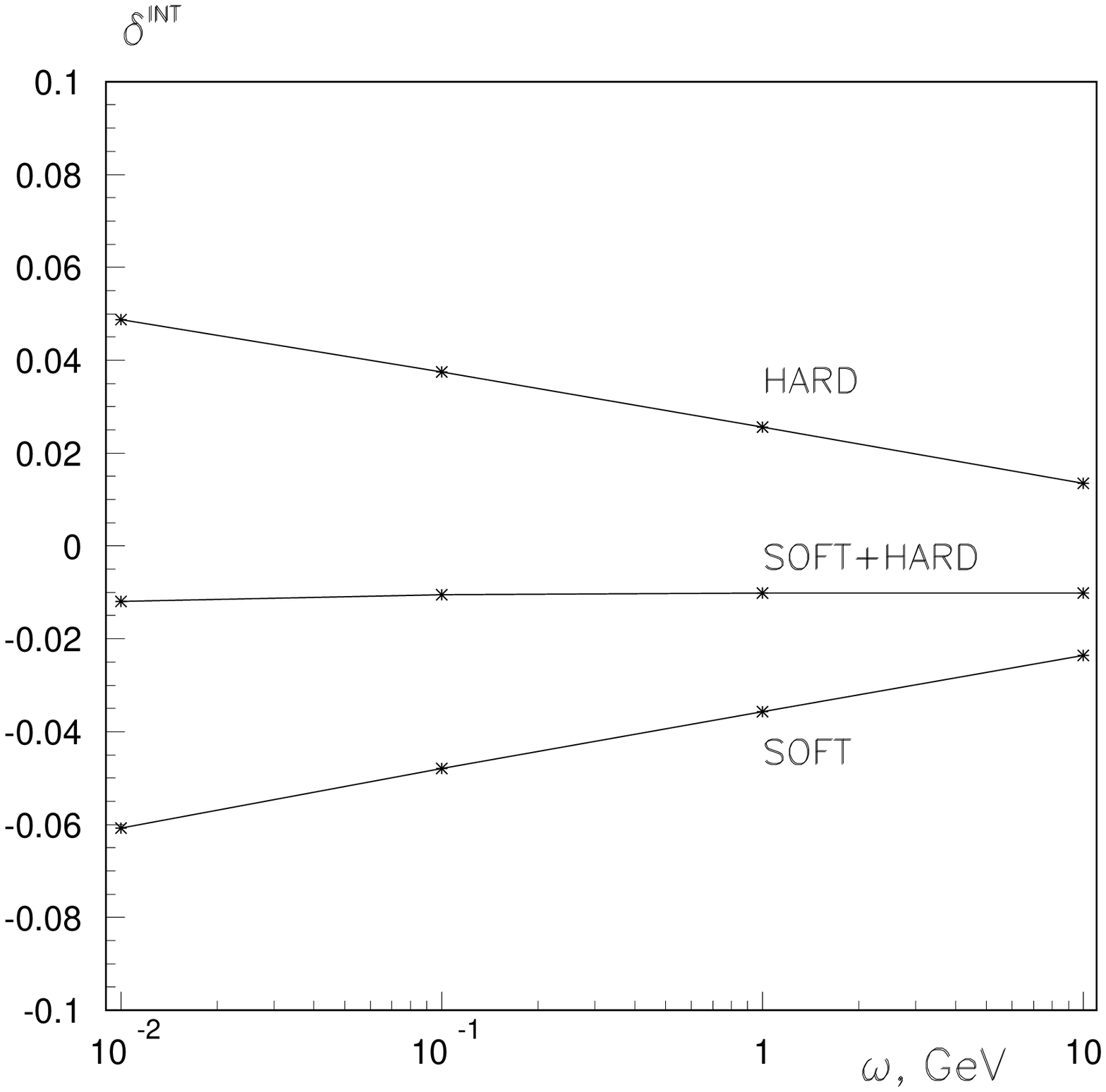}}
\vspace*{-5mm}
\caption{\label{fig:2}
Independence of the FSR, ISR and INT-parts of bremsstrahlung cross section
on separator $\omega$.
}
\end{figure*}

\begin{figure*}
\vspace*{45mm}
\hspace*{-30mm}
\scalebox{0.4}{\includegraphics{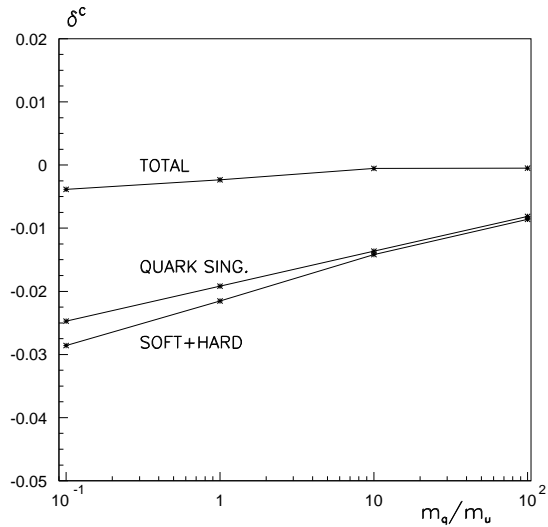}}
\vspace*{-15mm}
\caption{\label{fig:3}
Independence of the ISR-part of bremsstrahlung cross section
on quark mass $m_q$, the "TOTAL" means "SOFT+HARD-QUARK SING.",
we suppose here $\omega=10$ GeV.
}
\end{figure*}

First, we want to demonstrate the independence the results on
unphysical parameters: "soft"-"hard" photon separator 
$\omega$ (Fig.\ref{fig:2})
and the quark mass (Fig.\ref{fig:3}).
In these Figs. we can see the relative corrections
\begin{equation}
\delta^C = {\sigma^H_C}/{\sigma^H_0}
\end{equation}
to cross sections integrated over interval of invariant dimuon mass
1~TeV $\leq M \leq$ 14~TeV and assuming $\zeta^*=1$ and $p_T^{min}=0$.
Fig.\ref{fig:2} shows the $\omega$-independence for FSR (left picture), 
ISR (middle picture) and INT-part (right picture) separately in wide range
of $\omega$: $10^{-2}$~GeV~$\leq~\omega~\leq$~10~GeV.
We can see also the obvious property for sums the SOFT and the HARD parts:
$|\mbox{FSR}| > |\mbox{ISR}| > |\mbox{INT}|$; all of them are negative.

For the decision of the quark mass singularity problem we used
the $\rm \overline{MS}$ scheme \cite{MSbar}
and the procedure of linearization which is well-grounded in \cite{SANC}.
After all manipulations the part of cross section which
must be subtracted to be free from the quark mass dependence
has the form (here we used abbreviations Q.S.=QUARK SING. 
and $q(x) = q(x,Q^2) \equiv f_q(x,Q^2)$)
\begin{eqnarray}
\sigma_{Q.S.}^H =
\frac{1}{3}
\int\limits_{0}^{1}dx_1
\int\limits_{0}^{1}dx_2
\int\limits_{0}^{1}dz
\int\limits_{-S}^{0}dt
\sum\limits_{q=u,d,...}
&&
\left[ \left( q(x_1)\Delta \bar q(x_2,z) \theta(z-x_2) 
 +\Delta q(x_1,z)\bar q(x_2) \theta(z-x_1) )
\hat \sigma_{0}^{q\bar q}(t) \right.\right. +
 \nonumber \\[0.3cm] \displaystyle && 
\left.\left. + (q \leftrightarrow \bar q) \right) \right]
\theta(t+\hat s)
\theta_M
\hat \theta_{D},
\label{quarksing}
\end{eqnarray}
where
\begin{eqnarray}
\Delta q(x,z) =
\frac{\alpha}{2\pi}Q_q^2
\left[ \frac{1}{z}q(\frac{x}{z},M_{sc}^2)-q(x,M_{sc}^2)\right]
\frac{1+z^2}{1-z}
\left( \log\frac{M_{sc}^2}{m_q^2}-2\log(1-z)-1\right)
\label{deq}
\end{eqnarray}
and $M_{sc}$ is the factorization scale \cite{MSbar}.
Fig.\ref{fig:3} shows the $m_q$-independence for ISR part of cross section
(the asterisks on plot are the points where the calculation was made,
they are connected by straight lines).
We can see that in the range of rather big values (10 -- 100) 
of ratio $m_q/m_u$
the difference (SOFT+HARD)-(QUARK SING.) is constant
(i.e. independent on $m_q$).
In the region of small $m_q$ this property is slightly broken.
The reason is simple: at small parameter of mass the calculation
of mass singularity cross section demands of more time 
(it is better to say -- more iterations in integration). 
In Fig.\ref{fig:3} all of points are obtained with the same number of iterations,
so in the region of small $m_q$ the result for HARD ISR part has
not so good accuracy, in actual fact this part is slightly more.
Surely increasing the accuracy (and simultaneously the running time
of code) we provide the exact cancellation of $m_q$, 
this obvious graph of less importance than Fig.\ref{fig:3} and we do not present
it here.

\begin{figure*}
\vspace*{45mm}
\hspace*{-40mm}
\scalebox{0.4}{\includegraphics{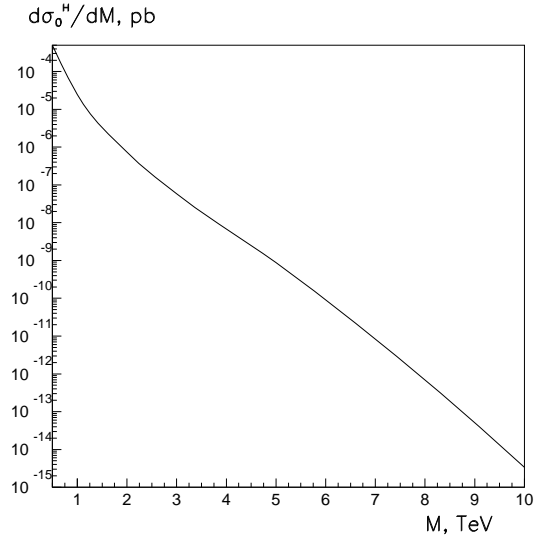}}
\vspace*{-15mm}
\caption{\label{xs0}
The differential Born cross section 
as a function of $M$.
}
\end{figure*}

\begin{figure*}
\vspace*{45mm}
\hspace*{-40mm}
\scalebox{0.4}{\includegraphics{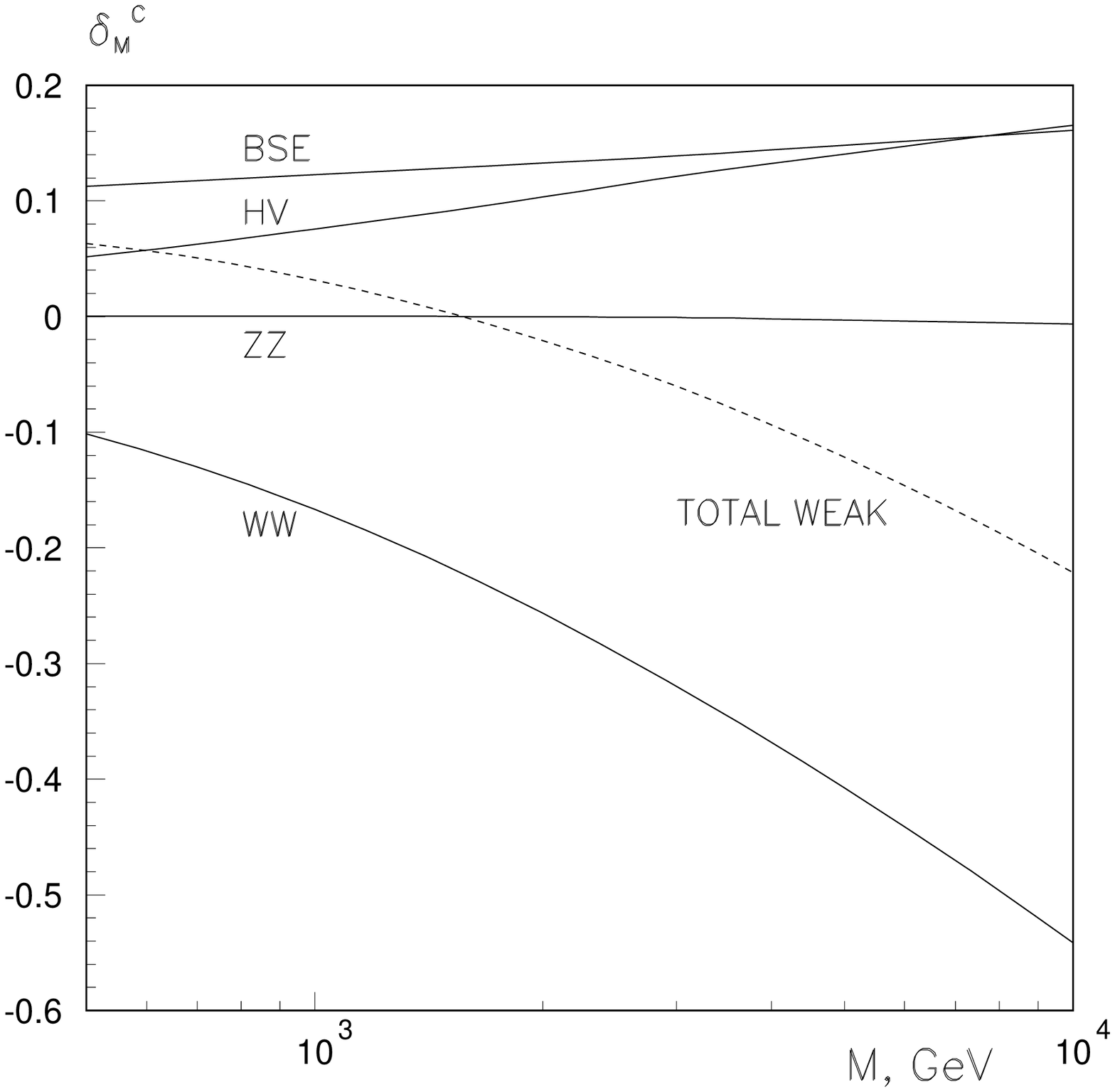}}
\hspace*{30mm}
\scalebox{0.4}{\includegraphics{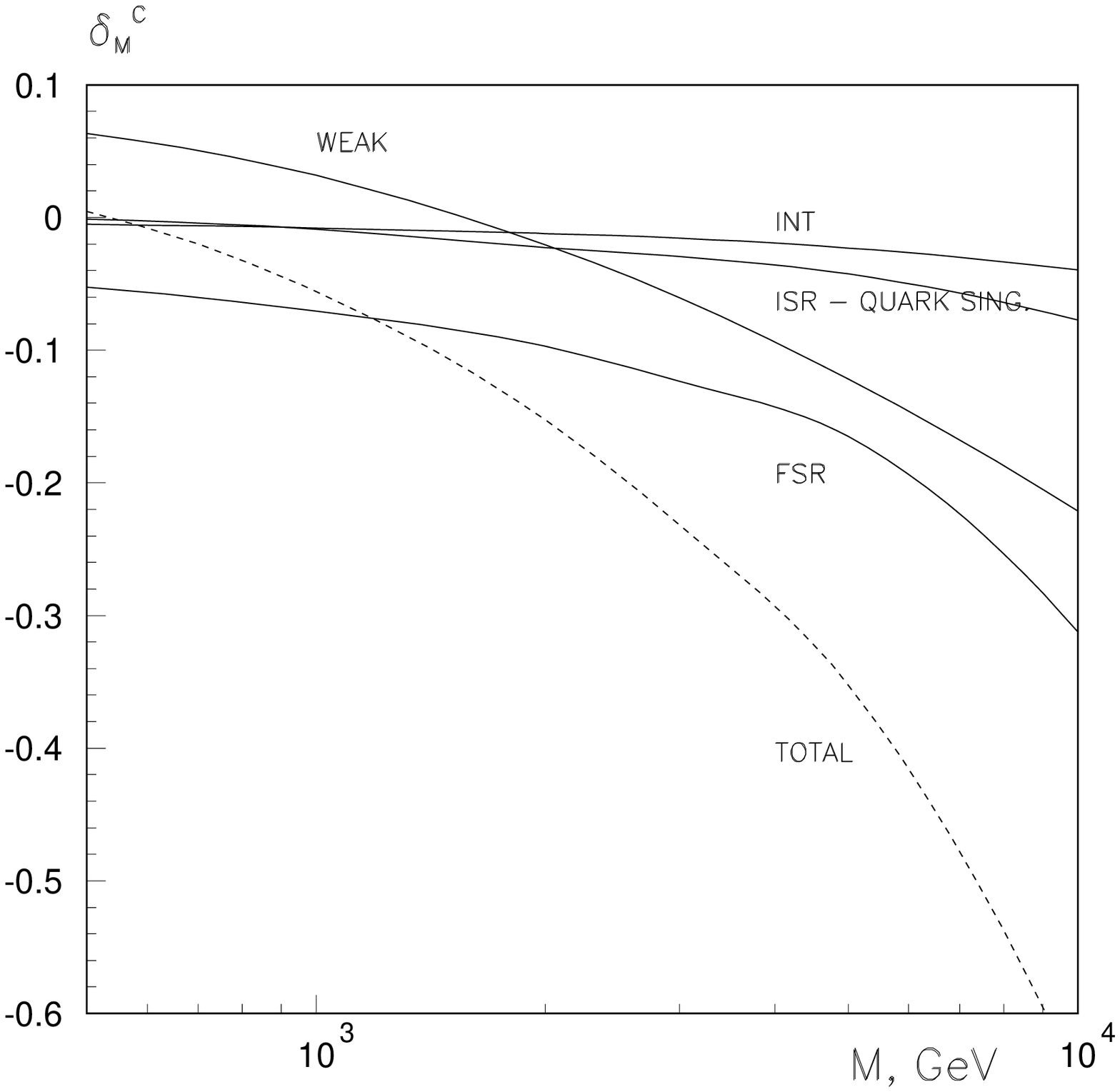}}
\vspace*{-15mm}
\caption{\label{RC}
The relative corrections $\delta_M^C$
(pure weak corrections -- left picture
and including the QED corrections -- right picture) as a functions of $M$.
}
\end{figure*}

In the following using FORTRAN program READY 
\footnote{FORTRAN code READY is available by contacting to author via e-mail}
(READY is "Radiative corrEctions to lArge invariant mass Drell-Yan process")
the scale of electroweak radiative corrections and their effect
on the observables of the Drell-Yan processes for future CMS experiments
will be discussed.
In Fig.\ref{xs0} and Fig.\ref{RC} it is shown 
the differential Born cross section 
and the relative corrections to it
\begin{equation}
\delta_M^C = \frac{d\sigma^H_C}{dM}/\frac{d\sigma^H_0}{dM}
\end{equation}
as a functions of $M$.
The pure weak (total electroweak) corrections are presented 
in left (right) picture of Fig.\ref{RC}. 
The translation from total to the differential cross sections
realized with the help of trick presented in Appendix~B.

We used the following set of prescriptions:
\begin{itemize}
\item
investigated reaction is (1) with 
the energy of LHC $\sqrt{S}=14\ \mbox{TeV}$,
\item
the set of SM input electroweak parameters:
$\alpha=1/137.03599911$,\ 
$m_Z=91.1876\ \mbox{GeV}$,\ $m_W=80.37399\ \mbox{GeV}$,\ 
$\Gamma_Z=2.4924\ \mbox{GeV}$,\ $\Gamma_W=2.0836\ \mbox{GeV}$,\ 
$m_H=115\ \mbox{GeV}$,
\item
muon mass $m_\mu=0.105658369\ \mbox{GeV}$,
masses of fermions for loop contributions to the BSE:
$m_e=0.51099892\ \mbox{keV}$,\ $m_\tau=1.77699\ \mbox{GeV}$,\ 
$m_u=0.06983\ \mbox{GeV}$,\ $m_c=1.2\ \mbox{GeV}$,\ $m_t=174\ \mbox{GeV}$,
$m_d=0.06984\ \mbox{GeV}$,\ $m_s=0.15\ \mbox{GeV}$,\ $m_b=4.6\ \mbox{GeV}$
(the light quark masses provide $\Delta \alpha_{had}^{(5)}(m_Z^2)$=0.0276),
\item
5 active flavors of quarks in proton, their masses as regulators
of the collinear singularity $m_q=10\times m_u$,
\item
non-diagonal elements of CKM matrix = 0, diagonal ones = 1,
\item
"soft"-"hard" photon separator $\omega=10$ GeV,
\item
the MRST2004QED set of unpolarized parton distribution functions 
\cite{MRST} (with the choice $Q=M_{sc}=m_Z$),
\item
we impose the experimental restriction conditions
on the detected lepton angle
$-\zeta^* \leq \zeta \leq \zeta^*$
and on the rapidity $|y(l)|\leq y(l)^*$, see (\ref{tetad});
for CMS detector the cut values of $\zeta^*$ and $y(l)^*$
are determined as
\begin{equation}
y(l)^* = - \ln \ \tan \frac{\theta^*}{2} = 2.5,\
\zeta^* = \cos\theta^* \approx 0.986614,
\label{restr}
\end{equation}
also we used the second standard CMS restriction
$p_T(l) \geq 20\ \mbox{GeV}$,
\item
here we used so-called "bare" setup for muons identification requirements
(no smearing, no recombination of muon and photon).
\end{itemize}

Let us discuss briefly the effects of EWC induced by different
contributions (in region $M=1~\mbox{TeV}$). 
The BSE-contribution is positive and $\sim 0.12$, it is usual effect of BSE.
The HV-part gives positive contribution $\sim 0.07$ 
in spite of the negative sigh of DSL 
($-\log^2({m_{Z(W)}^2/s})=-l^2_{Z(W),s}$)
in diagrams Fig.1,b and Fig.1,c with the Z and W
as additional virtual particle. 
Analysis shows that the Single Sudakov Logs (SSL=$l_{Z(W),s}$) 
of diagrams Fig.1,d and Fig.1,e play the very 
important role in the region of TeV's M.
To determine that we can compare, for example, 
the coefficients at the functions $\Lambda_2(k^2,M_W)$ 
(it contains DSL and SSL) and  $\Lambda_3(k^2,M_W)$ 
(it contains only SSL)
in formulas (6.8)-(6.12) from \cite{BSH86},
the second one is sometimes much more than first one (up to 9 times),
whereas $|\mbox{DSL/SSL}|=|\mbox{SSL}|\approx$4.79 at $M=1\ \mbox{TeV}$.
The combined effect of all HV becomes positive.
Then, the WW boxes are uniquely negative and they are the dominant 
contributions.
Let us explain it by the example of
WW-diagram and $\gamma$-exchange Born diagram interference:
extracting this part of cross section 
(denote it $\sigma^{u\bar u+\bar u u}_{WW \times \gamma}$)
and retaining only u-type of quark contributions and leading power
of Sudakov logs (there is no SSL in WW-boxes).
Then (see formula (40) of \cite{PRD})
\begin{equation}
\sigma^{u\bar u+\bar u u}_{WW \times \gamma} \sim
u \bar u \cdot \delta^{WW,\gamma}(t,u,b_+,b_-) +
\bar u u \cdot \delta^{WW,\gamma}(u,t,b_+,b_-),
\end{equation}
here $u (\bar u) \equiv f_{u(\bar u)}^p(x_{1(2)})$.
In that way we can see the fact: the terms $u\bar u$ and $\bar u u$
contain the same $b$ 
and different invariants $t$ and $u$ as arguments of $\delta^{WW,\gamma}$.
Further, using $b_+^{WW,\gamma} = -2{(v^{WW})}^2 Q_q$
and $b_-^{WW,\gamma}=0$
(see (\ref{bplusminus}), (\ref{lamb}),
and $v^{ij}=v^iv^j+a^ia^j$) we can make sure that
\begin{equation}
\sigma^{u\bar u+\bar u u}_{WW \times \gamma} \sim
-2{(v^{WW})}^2 Q_u [u \bar u \cdot t^2 l^2_{W,t} 
  + \bar u u\cdot u^2 l^2_{W,u} ] < 0.
\end{equation}
Corresponding contribution of d-type of quarks also less than zero and
looks like
\begin{equation}
\sigma^{d\bar d+\bar d d}_{WW \times \gamma} \sim
2{(v^{WW})}^2 Q_d [d \bar d \cdot t^2 l^2_{W,t} 
  + \bar d d\cdot u^2 l^2_{W,u} ] < 0.
\end{equation}
The same situation takes place also for ${WW \times Z}$-case.
At last, ZZ-, ISR-, INT- parts are small enough to give determinant effect
(ISR gives $\sim -0.019$, INT gives $\sim -0.008$, ZZ is $\sim 0.0003$)
and FSR-part is negative and $\sim -0.071$, so the total effect of EWC 
is found to be negative $\sim -0.056$.


\section{Conclusions}
The complete electroweak radiative ${\cal O}{(\alpha)}$ 
corrections to the Drell-Yan process
at large invariant dimuon mass have been studied.
For the shortening of code running time (keeping an enough accuracy)
we simplify the calculation as much as possible (using AA and 
generalized functions).
Using FORTRAN code READY the numerical analysis is performed 
in the high energy region corresponding to the future experiments
at the CERN Large Hadron Collider.
To simulate the detector acceptance we used the standard CMS detector cuts.
The radiative corrections are found to become large at 
high dimuon mass $M$, the complete corrections at "bare" setup change 
the dimuon mass distribution up to $\sim -5.6\%\ (-23.2\%;\ -35.3\%) $
at the LHC energy and $M=1\ (3;\ 5)\ \mbox{TeV}$.

Some issues have became beyond the scope
of the presented paper (the detailed numerical analysis of 
process $pp \rightarrow e^+e^-X$ 
and other interesting observables: total inclusive cross section,
forward-backward asymmetries; al last, "calo" results -- 
taking into consideration also smearing and recombination).
All that will be the subjects of future investigation
but, first of all, due the importance and complexity 
of investigated problem, 
we should cross-check with the results of other groups 
(programs SANC \cite{SANC}, ZGRAD \cite{DY2002, ZGRAD}, ...)
to make sure that our result is correct.
Author will be grateful to all interested groups for 
giving a chance to compare the results in that stage.

\section{Acknowledgments}
I am grateful to 
A.~Arbuzov, D.~Bardin, S.~Bondarenko, I.~Golutvin, 
A.~Ilyichev, E.~Kuraev, A.~Lanyov,
V.~Mossolov, S.~Shmatov, N.~Shumeiko, D.~Wackeroth 
for the stimulating discussions.

\appendix

\section{Expressions for the $R$}

The expressions for the $R$ have such form:
for lepton emission (see Fig.1,i and Fig.1,j)
\begin{eqnarray}
{R_l}_+^{q\bar q}=&&
 2  s
-2 \frac{m^2}{z^2}
(2 u_1^2 + 2 u_1 s + 4 u_1 t - 4 u_1 v + s^2 + 2 s t
  -2 s v + 2 t^2 - 4 t v + 2 v^2)
\nonumber \\[0.3cm] \displaystyle
&&- 2 \frac{s}{zv}
( - u_1^2 - u_1 s - 2 u_1 t - s^2 - 2 s t - 2 t^2)
- \frac{s}{z}
(4 u_1 + 4 s + 6 t - 3 v)
\nonumber \\[0.3cm] \displaystyle
&&- 2 \frac{m^2}{v^2}
(s^2 + 2 s t + 2 t^2)
- \frac{1}{v}
s ( - z + 2 u_1 + 2 s + 2 t),
\nonumber \\[0.3cm] \displaystyle
{R_l}_-^{q\bar q}=&&
2 \frac{m^2}{z^2}
s (2 u_1 + s + 2 t - 2 v)
+ 2 \frac{s^2}{zv}
( - u_1 - s - 2 t)
 + \frac{s}{z}
(4 s + 2 t - v)
\nonumber \\[0.3cm] \displaystyle
&&
+ 2 \frac{m^2}{v^2}
s (s + 2 t)
+ \frac{s}{v}
( - z + 2 u_1 + 2 s + 2 t) - 2 s,
\end{eqnarray}
for quark emission (see Fig.1,k and Fig.1,l)
\begin{eqnarray}
{R_{qk}}_+^{q\bar q}=&&
 2 ( z - s + v)
- 2 \frac{m_q^2}{z_1^2}  (z^2 + 2 z t + s^2 + 2 s t - 2 s v + 2 t^2 - 2 t v + v^2)
\nonumber \\[0.3cm] \displaystyle &&
- 2 \frac{s}{z_1 u_1} ( - s^2 - 2 s t + s v - 2 t^2 + 2 t v - v^2)
\nonumber \\[0.3cm] \displaystyle &&
- \frac{1}{z_1} (z^2 - z s + 2 z t + 2 s^2 + 2 s t - s v + 4 t^2 - 2 t v + v^2)
\nonumber \\[0.3cm] \displaystyle &&
- 2 \frac{m_q^2}{u_1^2}  (z^2 - 2 z s - 2 z t + 2 z v + s^2 + 2 s t
	- 2 s v + 2 t^2 - 2 t v + v^2)
\nonumber \\[0.3cm] \displaystyle &&
- \frac{1}{u_1} (z^2 - 3 z s - 2 z t + 2 z v + 4 s^2 + 6 s t - 5 s v
	+ 4 t^2 - 6 t v + 3 v^2),
\nonumber \\[0.3cm] \displaystyle
{R_{qk}}_-^{q\bar q}=&&
2 \frac{m_q^2}{z_1^2}  ( - z^2 - 2 z t + s^2 + 2 s t - 2 s v - 2 t v + v^2)
+ 2 \frac{ s^2}{z_1 u_1} ( - s - 2 t + v)
\nonumber \\[0.3cm] \displaystyle &&
+ \frac{1}{z_1} ( - z^2 + z s - 2 z t + 2 s^2 + 6 s t - 3 s v - 2 t v + v^2)
\nonumber \\[0.3cm] \displaystyle &&
+ 2 \frac{ m_q^2}{u_1^2} (z^2 - 2 z s - 2 z t + 2 z v + s^2 + 2 s t - 2 s v
	- 2 t v + v^2)
\nonumber \\[0.3cm] \displaystyle &&
+ \frac{1}{u_1} (z^2 - 3 z s - 2 z t + 2 z v + 4 s^2 + 6 s t - 5 s v - 2 t v + v^2),
\end{eqnarray}
for lepton-quark interference 
\begin{eqnarray}
{R_{int}}_+^{q\bar q}=&&
  2 (z - u_1 - s - 4 t + 3 v)
+ \frac{t}{z z_1} (2 s^2 + 4 s t - 2 s v + 4 t^2 - 2 t v + v^2)
\nonumber \\[0.3cm] \displaystyle
&&
+\frac{1}{zu_1} (2 s^3 + 6 s^2 t - 6 s^2 v + 8 s t^2 - 14 s t v
        + 7 s v^2 + 4 t^3 - 10 t^2 v + 9 t v^2 - 3 v^3)
\nonumber \\[0.3cm] \displaystyle
&&
+ \frac{2}{z} (u_1 s - u_1 v + s^2 + 2 s t - 3 s v - 3 t v + 2 v^2)
\nonumber \\[0.3cm] \displaystyle
&&
+ \frac{1}{z_1v} (z^2 s + z^2 t + 2 z s t + 2 z t^2 + 2 s^3 + 6 s^2 t
        + 8 s t^2 + 4 t^3)
+ \frac{1}{z_1} (z t - 2 s^2 - 4 s t + s v + t v)
\nonumber \\[0.3cm] \displaystyle
&&
+ \frac{t}{v u_1} (z^2 - 2 z s - 2 z t + 2 s^2 + 4 s t + 4 t^2)
+ \frac{2}{v} ( - z t - u_1 s - s^2 - 2 s t)
\nonumber \\[0.3cm] \displaystyle
&&
+ \frac{1}{u_1} ( - z^2 + 3 z s + 5 z t - 3 z v - 4 s^2 - 12 s t + 8 s v
        - 12 t^2 + 13 t v - 5 v^2),
\nonumber \\[0.3cm] \displaystyle
{R_{int}}_-^{q\bar q}=&&
 2 (-z + 2 s - v)
- \frac{t}{z z_1} (2 s^2 + 4 s t - 2 s v - 2 t v + v^2)
\nonumber \\[0.3cm] \displaystyle
&&
- \frac{1}{z u_1} (2 s^3 + 6 s^2 t - 6 s^2 v + 4 s t^2 - 10 s t v
       + 5 s v^2 - 2 t^2 v + 3 t v^2 - v^3)
\nonumber \\[0.3cm] \displaystyle
&&
- \frac{2s}{z} (s - v)
- \frac{1}{z_1 v} ( - z^2 s - z^2 t - 2 z s t - 2 z t^2 + 2 s^3
       + 6 s^2 t + 4 s t^2)
\nonumber \\[0.3cm] \displaystyle
&&
- \frac{1}{z_1} ( - z t - 2 s^2 - 4 s t + s v - 4 t^2 + t v)
- \frac{t}{v u_1} (z^2 - 2 z s - 2 z t + 2 s^2 + 4 s t)
\nonumber \\[0.3cm] \displaystyle
&&
- \frac{2s}{v} (z - s)
- \frac{1}{u_1} ( - z^2 + 3 z s + 5 z t - 3 z v - 4 s^2 - 12 s t + 8 s v
       - 4 t^2 + 7 t v - 3 v^2).
\end{eqnarray}

\section{Translation from total to the differential cross section}

To translate the total non-radiative  (\ref{xsfin}) 
(or radiative (\ref{xshard})) cross section 
to the differential one we should just differentiate it 
on variable $M$ using obvious rule
\begin{equation}
\frac{d\sigma}{dM}=-\left.{\frac{d\sigma}{dM_1}}\right|_{M_1=M}.
\end{equation}
After that and taking into consideration
the formula 
$$\frac{d\theta(x)}{dx}=\delta(x) $$
we can significantly simplify the form of cross section, as
we are in a position to integrate analytically over one of variable
(here we choose $x_2$) in such a way
$$  \int_a^b f(x)\delta(x-z) dx= f(z)\theta(z-a)\theta(b-z).$$
Finally, we get very simple recipe for translation from
total to the differential cross section:
so, for radiative case  to pass 
$$\sigma^H_{hard}  \rightarrow \frac{d\sigma^H_{hard}}{dM}$$ 
we should in formula (\ref{xshard})
\begin{enumerate}
\item do not integrate over $x_2$ and omit $\theta(\hat s-z-v-M_1^2)$,
\item substitute $x_2 \rightarrow {(M^2+z+v)}{(Sx_1)^{-1}}$ 
(or $\hat s  \rightarrow M^2+z+v$),
\item multiply by factor ${2M}{(Sx_1)^{-1}}\theta(Sx_1-M^2-z-v)$.
\end{enumerate}
For nonradiative case the translation steps are the same but 
"radiative" invariants should equal to zeros: $v = z = 0$ since
in that case $p \rightarrow 0$.


\begin {thebibliography}{99}
\bibitem {34} M.~Cveti{$\rm \check c$} and S.~Godfrey, hep-ph/9504216;
   T.G.~Rizzo, Proceedings of the 1996 DPF/DPB Summer Study on New Directions 
   for High Energy Physics Snowmass96, Snowmass, CO, 25 June - 12 July (1996);
   J.L.~Hewett and T.G.~Rizzo, Phys. Rept. 183 (1989) 193.
\bibitem {extra-dim} N.~Arkani-Hamed et.al, Phys. Lett. B 
   {\bf 429}, 263 (1998) [arXiv:hep-ph/9803315];
   I.~Antoniadis et.al, Phys. Lett B {\bf 436}, 
   257 (1998) [arXiv:hep-ph/9804398];
   L.~Randall and R.~Sundrum, Phys. Rev. Lett. {\bf 83}, 3370 (1999) 
   [arXiv:hep-ph/9905221], \ Phys. Rev. Lett. {\bf 83}, 4690 (1999) 
   [arXiv:hep-th/9906064];
   C.~Kokorelis, Nucl. Phys. {\bf 677} (2004) 115 [arXiv:hep-th/0207234]
\bibitem {extra-bos} A.~Leike, Phys. Rep. {\bf 317}, 143 (1999), 
   [arXiv:hep-ph/9805494];
   T.G. Rizzo, arXiv:hep-ph/9612440
\bibitem {89} D.~Bourilkov, arXiv:hep-ph/0305125; CERN-CMS-NOTE-2006-085 (2006)
\bibitem {cmsnote}
   I.~Belotelov et.al, CERN-CMS-NOTE-2006-123 (2006)
\bibitem {MosShuSor} V.~Mosolov, N.~Shumeiko, Nucl.Phus. B {\bf 186},
   397 (1981);  A.~Soroko, N.~Shumeiko, Yad. Fiz {\bf 52}, 514 (1990)
\bibitem {DY2002} U.~Baur et al.,  Phys. Rev. D {\bf 65}: 033007, (2002)
   [arXiv:hep-ph/0108274]
\bibitem {sud-log} V.~Sudakov, Sov. Phys. JETP {\bf 3}, 65 (1956)
\bibitem {DENPOZ} A.~Denner and S.~Pozzorini,
   Eur.\ Phys.\ J.\ C {\bf 18} (2001) 461 [arXiv:hep-ph/0010201];\ 
   Eur.\ Phys.\ J.\ C {\bf 21} (2001) 63 [arXiv:hep-ph/0104127]
\bibitem {ARX05} B.~Jantzen, J.H.~Kuhn, A.A.~Penin, and V.A.~Smirnov,
   TTP05-17, PSI-PR-05-04 [arXiv:hep-ph/0509157]
\bibitem {ZGRAD} \verb|http://ubhex.physics.buffalo.edu/~baur/zgrad2.tar.gz|
\bibitem {Baur2006}  U.~Baur,  Phys.\ Rev.\ D {\bf 75}, 013005 (2007)
   [arXiv:hep-ph/0611241]
\bibitem{Denner:2006jr} A.~Denner, B.~Jantzen and S.~Pozzorini,
   Nucl.\ Phys.\  B {\bf 761} (2007) 1 [arXiv:hep-ph/0608326]
\bibitem{CarCal} C.M.~Carloni Calame {\it et al.} JHEP {\bf 0505}, 019 (2005) 
\bibitem {YAFDY} V.~Zykunov, Yad. Fiz. {\bf 69}, 1557 (2006) 
   (Engl. vers.: Phys. of Atom. Nucl. {\bf 69}, 1522 (2006))
\bibitem {PRD} V.~Zykunov,  Phys.\ Rev.\ D {\bf 75}, 073019 (2007) 
    [arXiv:hep-ph/0509315]
\bibitem {BSH86}  M.~B\"ohm, H.~Spiesberger, W.~Hollik, Fortschr. Phys. 
   {\bf 34}, 687 (1986)
\bibitem {Hollik}  W.~Hollik,  Fortschr. Phys. {\bf 38}, 165 (1990)
\bibitem {baur} U.~Baur {\it et al.},  Phys. Rev. D {\bf 57}, 199 (1998)
\bibitem {VEGAS} G.~Peter Lepage, J. Comput.Phys. {\bf 27}, 192 (1978)
\bibitem {MSbar} W.~Bardeen {\it et al.},  Phys. Rev. D {\bf 18}, 3998 (1978)
\bibitem {SANC} A.~Andonov, A.~Arbuzov, D.~Bardin et al.,
   Comput. Phys. Commun. {\bf 174}, 481 (2006) [arXiv:hep-ph/0411186];\
   SANC project website: \verb|http://sanc.jinr.ru, http://pcphsanc.cern.ch|
\bibitem {MRST} A.D.~Martin {\it et al.} Eur.Phys.J. C {\bf 39}, 155 (2005)
   [arXiv:hep-ph/0411040]
\end {thebibliography}
\end{document}